\documentclass[fleqn,10pt]{wlscirep}
\usepackage[utf8]{inputenc}
\usepackage[T1]{fontenc}
\title{Spontaneous center formation in Dictyostelium discoideum}

\author[1,*]{Estefania Vidal-Henriquez}
\author[1,+]{Azam Gholami}
\affil[1]{Max Planck Institute for Dynamics and Self-Organization, Am Fassberg 17, D-37077 G{\"o}ttingen, Germany}

\affil[*]{estefania.vidal@ds.mpg.de }

\affil[+]{azam.gholami@ds.mpg.de}

\begin{abstract}
\textit{Dictyostelium discoideum} (\textit{D.d.}) is a widely studied amoeba due to its capabilities of development, survival, and self-organization. During aggregation it produces and relays a chemical signal (cAMP) which shows spirals and target centers. Nevertheless, the natural emergence of these structures is still not well understood. We present a mechanism for creation of centers and target waves of cAMP in \textit{D.d.} by adding cell inhomogeneity to a well known reaction-diffusion model of cAMP waves and we characterize its properties. We show how stable activity centers appear spontaneously in areas of higher cell density with the oscillation frequency of these centers depending on their density. The cAMP waves have the characteristic dispersion relation of trigger waves and a velocity which increases with cell density.
Chemotactically competent cells react to these waves and create aggregation streams even with very simple movement rules. Finally we argue in favor of the existence of bounded phosphodiesterase to maintain the wave properties once small cell clusters appear.  
\end{abstract}
\begin{document}

\flushbottom
\maketitle
%
%
\thispagestyle{empty}


\section*{Introduction}

\textit{Dictyostelium discoideum} (\textit{D.d.}) is a social amoeba that under adverse environmental conditions begins a developmental program in order to survive\cite{kessin2001dictyostelium}. After 4-5 hours into this program, the cells start to produce the signaling chemical cAMP (cyclic adenosine monophosphate), which travels the system as a wave relayed by the cells. The cells are chemotactically competent and move against the traveling wave of cAMP towards the aggregating centers. In the centers the cells form a mound, where they further develop and differentiate into a slug and eventually into a fruiting body who releases spores which can become myxamoebas; thus completing the life cycle of the amoeba.

The chemical signaling part of the process presents spirals and target centers, these structures are characteristic of both oscillatory and excitable systems, and are present in many systems in nature such as the cardiac muscle \cite{allessie1973circus}, calcium waves in oocytes \cite{Lechleiter123}, NADH waves in glycolysis \cite{mair1996traveling}, among others. Other important system where such structures are observed is the chemical reaction known as Belousov-Zhabotinsky \cite{winfree1972spiral,zaikin1970concentration}, which has become a model experimental system to study spirals and target waves.

A distributed oscillatory system would produce bulk oscillations if it is well stirred, that is, the whole system oscillates synchronously. If it is not stirred and presents some inhomogeneities, such as scratches on the container surface or dust particles, concentric circular waves will appear \cite{kopell1973plane,stich2002complex}. We refer to these structures as target patterns. Once a wave breaks, it creates a spiral wave, a persistent structure with topological charge\cite{keener1986geometrical,tyson1988singular}.

Many models have been proposed to describe the cAMP waves in \textit{D. d.}\cite{levine1996positive,noorbakhsh2015modeling,martiel1987model}, from qualitative excitable models to models derived from the chemical reactions of production of cAMP from ATP. Each model has their own particular way of breaking the homogeneity and producing waves. Here we present the properties of the model proposed by Martiel and Goldbeter \cite{martiel1987model,tyson1989spiral} when cell inhomogeneity is introduced. We show through numerical simulations that cell inhomogeneity produces naturally stable target patterns which are centered in areas of higher cell density and relayed in the areas of lower density. Therefore, the system behaves both as oscillatory and excitable depending on local density. The produced waves are shown to be of the trigger kind, in contrast to phase waves (or pseudo-waves, without chemical transport), through their characteristic dispersion relation \cite{aliev1994dynamics}. We also show that wave speed is density dependent. It has been shown in other chemotaxis models that this dependence is enough to produce the wave instability responsible for cell streaming once cell movement is added\cite{hofer1997streaming,van1996spatial}. Our model reproduces that characteristic streaming pattern of this biological system.
Finally we show that a form of degradation which scales with local cell density is needed to have streaming at higher local densities, this degradation can exist in the form of phosphodiesterase bounded to the cell membrane.  

\section*{Results}

Our work combines a cellular automata scheme with the model proposed by Martiel and Goldbeter \cite{martiel1987model} and extended by Tyson \cite{tyson1989spiral} where three fields are described. The extracellular concentration of cAMP is represented as $\gamma$, the intracellular concentration as $\beta$, and $\rho$ represents the percentage of active receptors on the cell membrane. The receptors state changes between an active and an inactive state depending on the cAMP concentration at which they are exposed through the functions $f_1$ and $f_2$. The intracellular concentration of cAMP $\beta$ increases through production $\Phi(\rho,\gamma)$ and decreases through internal degradation $k_i$ and transport towards the extracellular media $k_t$. Finally the external concentration of cAMP can diffuse through the system, gets degraded by phosphodiesterase $k_e$ and increases due to transport from the internal media $k_t$.

We created a grid in our 1-D or 2-D system, dividing it in segments of size $r=10~\mu$m (or respectively squares of area $r^2$) and assigned random locations for the cells, each cell had their own value for the intracellular and membrane bound variables $\rho_i$ and $\beta_i$, while $\gamma(x,y)$ was assigned to a square in space. The used equations look as follows    

\begin{subequations}
	\begin{align}
	k_1^{-1}\partial_t\rho_i&=-f_1(\gamma(x_i,y_i))\rho_i+f_2(\gamma(x_i,y_i))(1-\rho_i),\\
	\partial_t\beta_i&=s\Phi(\rho_i,\gamma(x_i,y_i))-(k_i+k_t)\beta_i ,\\ 
	\partial_t\gamma(x,y)&=D\nabla^2\gamma(x,y)-k_e\gamma(x,y) + \sum_i^N H(i,x,y) k_t\beta_i/h,
	\end{align}
	\label{e:MartielGoldbeter_3Comp}
\end{subequations}
\begin{equation*}
\centering
\begin{array}{cccccc}
\text{with}&f_1(\gamma)=\dfrac{1+\kappa\gamma}{1+\gamma},  &f_2(\gamma)=\dfrac{\mathcal{L}_1+\kappa\mathcal{L}_2c\gamma}{1+c\gamma},  &\Phi(\rho,\gamma)=\dfrac{\lambda_1+Y^2}{\lambda_2+Y^2},& \text{    and} 
&Y(\gamma,\rho)=\dfrac{\rho\gamma}{1+\gamma},
\end{array}
\end{equation*}

\noindent where $s=q\sigma\alpha/(1+\alpha)$ measures the production intensity, $(x_i,y_i)$ correspond to the Cartesian coordinates of the $i$-th cell, and $H$ is an indexing function such that $H(i,x,y)=1$ if $x_i\in(x-r/2,x+r/2)$ and $y_i\in(y-r/2,y+r/2)$ and $0$ otherwise.
In this way $\gamma$ will increase on a grid space if there is a cell producing cAMP in that location and only diffuse and be degraded if there are no cells in that space. All used parameter are selected following Lauzeral et al. \cite{lauzeral1997desynchronization} for their good agreement with experimental measures and are indicated in Table \ref{t:Parameters}. We keep $k_e$ the external media degradation rate as our control parameter.

The original homogeneous system studied by Tyson et al. \cite{tyson1989spiral} presented different regimes as $k_e$ was increased. In our set of parameters for low $k_e$ the system has one steady state which is stable. At $k_e^*\approx 4.3$ min\textsuperscript{-1} the system undergoes a Hopf bifurcation and a stable limit cycle appears. This is the oscillatory regime of the system. Upon further increasing $k_e$ two new steady states appear through a saddle-node bifurcation at $k_e^\dagger\approx 7.74$ min\textsuperscript{-1}, an unstable one and a stable but excitable one. At approximate the same time the limit cycle destabilizes, resulting in the excitable regime of the system. We set our parameter $k_e$ so that the system is in the oscillatory regime. For a detailed description of the different regimes present in this system refer to our previous work \cite{vidal2017convective}.


\subsection*{Oscillatory clusters}
\label{SS:Osc_Cluseters}
\begin{figure}[!h]
	\centering
	{\includegraphics[width=\columnwidth]{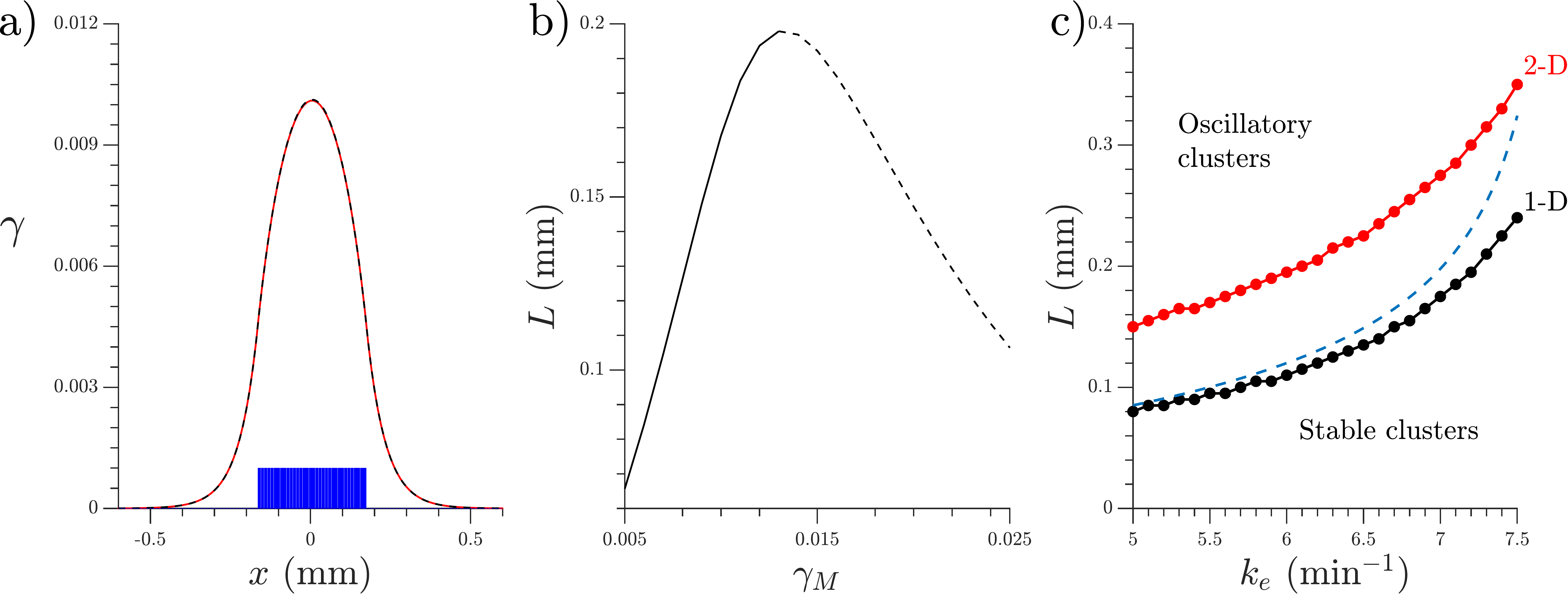}}
	\caption{ a) Time independent cluster size solution. Spaces occupied by cells in blue, result of numerical simulations in red, and theoretical prediction in black, dashed line. $k_e=7.0$ min \textsuperscript{-1} and $L=0.17$ mm, equivalent to $34$ consecutive cells. b) Distance $L$ from the center at which the system matches the decaying tail boundary condition, depending on the maximum concentration $\gamma_M$ reached at $x=0$. Stable solution in continuous line and unstable in dashed line. $k_e=7.0$ min \textsuperscript{-1}. c) Maximum size at which cluster still remain non-oscillatory. 1-D simulations in black, 2-D in red, theoretical 1-D prediction in dashed line.}
	\label{F:Stactic_Cluster}
\end{figure}

Numerical simulations of a small cluster of very closely located cells producing cAMP surrounded by buffer media without cells, reach a stable steady (non oscillatory) state due to cAMP diffusion to its surroundings. This state is shown in Fig. \ref{F:Stactic_Cluster} a). To approximate this solution we calculate separately the area with cells and the area without them. In the area without cells the system reduces to 
\begin{equation*}
\partial_t\gamma=D\partial_{xx}\gamma -k_e\gamma,
\end{equation*}
which in 1-D has a time independent decaying tail as solution. If the cluster size is $2L$, with the cells located in $x\in(-L,L)$ the decaying tail takes the form
\begin{equation*}
\gamma=Ae^{-\sqrt{k_e/D}|x|} \hspace{25 pt} \text{if } |x|>L
\end{equation*}
where $A$ is chosen to fulfill the boundary condition at $x=\pm L$, which are continuity, $\gamma(\pm L^-)$=$\gamma(\pm L^+)$, and continuity of the derivative, $\partial_x\gamma(\pm L^-)$=$\partial_x\gamma(\pm L^+)$. 

For the cAMP concentration inside the cell cluster we used the fact that simulation results showed that the $\gamma$ values inside the cluster are small compared to the values of the cAMP waves, which justifies making an approximation of the production function $\Phi$ for small values of $\gamma$, we therefore take the polynomial approximation $k_t\beta/h\approx a_0+a_1\gamma+a_2\gamma^2$ and the system reduces inside the cluster to
\begin{equation}
	0=a_0+(a_1-k_e)\gamma+a_2\gamma^2+D\partial_{xx}\gamma.
	\label{E:Approximated_production}
\end{equation}

It is worth mentioning that the polynomial approximation values were not taken from the Taylor expansion of the production function for $\gamma\ll1$, but rather from a least square fitting of the production function. The fitted values for our set of parameters (independent of $k_e$) are listed in Table \ref{Table_SystemParameters}. Equation (\ref{E:Approximated_production}) just by itself is invariant to space translation, even though this is not true for the entire system, this invariance produces a conserved quantity, which we will call energy and  can be used to calculate the shape of the cAMP accumulated around the cluster. We can rearrange the equation to show its energy conservation more clearly
\begin{equation*}
\partial_{xx}\gamma=-\dfrac{\partial V}{\partial\gamma} \hspace{15 pt} \text{with} \hspace{15 pt} V=\dfrac{a_0}{D}\gamma+\dfrac{(a_1-k_e)}{2D}\gamma^2+\dfrac{a_2}{3D}\gamma^3
\end{equation*}
and the total energy $E$ can be found by integrating both sides by $\gamma$. Therefore $E=V+(\partial_x\gamma)^2/2$ is conserved and fixed by the boundary condition. To calculate the shape of the solution we need the energy value, but since the decaying tail has a free parameter we need to introduce a second free parameter and then match the two solutions. Therefore, we leave the maximum value of $\gamma$, $\gamma_M$ as a free parameter and we numerically calculated the solution using

\begin{equation*}
\int_{\gamma}^{\gamma_M}\dfrac{d\gamma}{\sqrt{2[E(\gamma_M)-V(\gamma)]}}=x.
\end{equation*}
For every $\gamma_M$ there is a size $L$ such that it fits the boundary condition to match the decaying tail
\begin{equation*}
\sqrt{\dfrac{k_e}{D}}\gamma(x=L)=\partial_x\gamma|_{x=L}=\sqrt{2(E-V(x=L))},
\end{equation*}
an example of the relation $\gamma_M$ vs $L$ is shown in Fig. \ref{F:Stactic_Cluster} b). From there, it can be seen that it exists a maximum length for which the static solution exists. For those values of $L$ where two possible values of $\gamma_M$ exist, the system chooses the one with smaller $\gamma_M$. Indeed linear stability analysis showed that the smaller solution is stable, while the bigger one is unstable. Following this procedure we arrive at two results. First the cluster size values at which a stable, time independent solution exists, and second an approximation of this solution when it exists, see for comparison Fig. \ref{F:Stactic_Cluster} a) where both the numerical solution and theoretical prediction are plotted with excellent agreement.

The predicted maximum cluster size for different values is compared to the biggest clusters found through numerical simulations in one and two dimensions in Fig. \ref{F:Stactic_Cluster} c) with good agreement for small $k_e$. We believe the difference in agreement for large $k_e$ comes from the potential $V$ approximation, which is valid only for low values of $\gamma$, since for larger $k_e$ the values of $\gamma$ are higher, the approximation is not so good, but still manages to catch the general behavior of the system. For 2-D simulations, cells where located in $r<L$ with $r^2=x^2+y^2$. We see that bigger clusters can be maintained in a low steady state when the degradation is increased, this is expected since for fixed $L$, $\gamma_M$ diminishes with increasing $k_e$. 

If the cell group is bigger than its critical size it oscillates at a frequency that is size dependent, see for example Fig. \ref{F:Oscillating} a), also depicted in Supplementary Video S1, where one oscillating cluster is shown. As the cell group becomes larger, the frequency increases converging towards the limit cycle frequency. We measured this value for different cluster sizes once the cluster frequency converged to measuring precision. This relation is shown in Fig. \ref{F:Oscillating} d). 

At small sizes the cluster oscillation is synchronized (bulk), but as it gets bigger ($L\gtrapprox 0.19$ mm) a wave develops. This wave starts at the center of the cluster and moves towards the sides increasing its amplitude as it travels, as can be seen in Fig. \ref{F:Oscillating} b) and c). Compare to Fig. \ref{F:Oscillating} a) which shows a smaller cluster which oscillates synchronously. This effect can also be seen in Supplementary Video S2 which in turn can be compared to Supplementary Video S1.

\begin{figure}[!h]
	\centering
	{\includegraphics[width=\columnwidth]{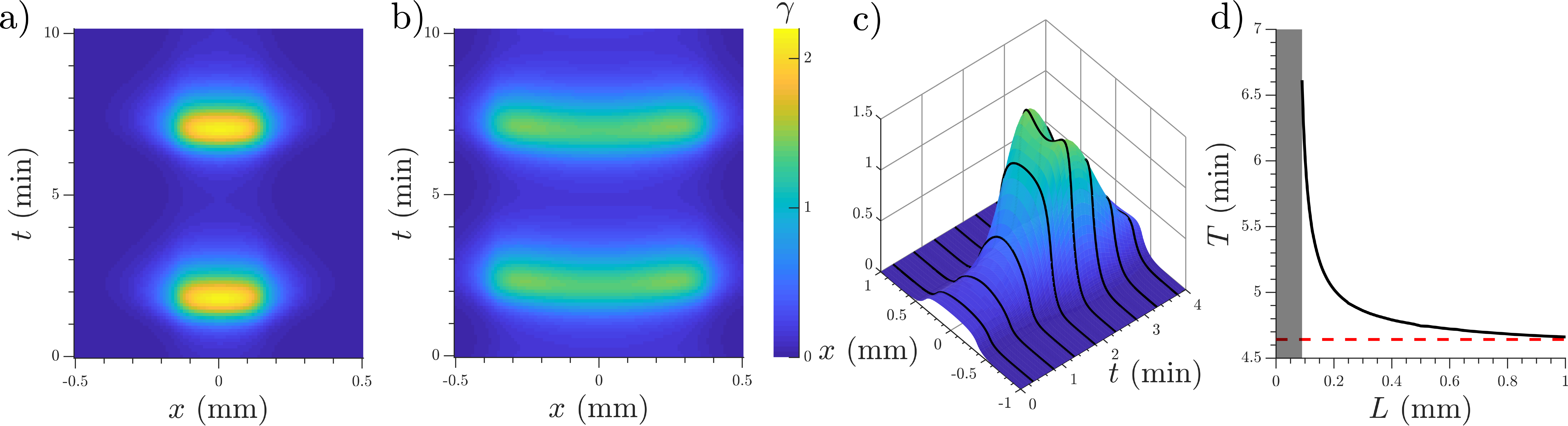}}
	\caption{ a) Space-time visualization of a self-oscillating cluster, with cells in $|x|<0.15$ mm. $k_e=5.0$ min\textsuperscript{-1}. b) A larger cluster of cells with $L=0.375$ mm and $k_e=5.0$ min\textsuperscript{-1}. c) Different representation of the simulation in b), concentration of cAMP is shown in the $z$ axis with equitemporal lines to guide the view.  The increase in amplitude towards the cluster edges can be observed (see also Supplementary Video S2). d) Oscillation period of isolated clusters depending on their size in black line. Cells located in $|x|<L$ and rest of the system empty space, $k_e=5.0$ min\textsuperscript{-1}. Red dashed line marks the period of the limit cycle in spaceless (0-D) simulations.}
	\label{F:Oscillating}
\end{figure}

\subsection*{Target patterns in random distributions of cells}
When we assigned random locations to the cells, the system showed different regimes as we increased $k_e$. For low values of $k_e$ the system oscillated mostly synchronously, see Supplementary Video S3 and Fig. \ref{F:OneD_Spacetimes} a) for a space-time plot. For high values of $k_e$ the system was incapable of oscillating by itself and reached a steady state of low cAMP (see Fig. \ref{F:OneD_Spacetimes} c)).

A more interesting behaviour was observed for intermediate values of $k_e$. There, the areas of higher local density acted as oscillatory centers, while the rest of the system relayed the emitted waves. This means that the lower density areas although not capable of oscillating by themselves are capable of producing enough cAMP to maintain the wave and avoid its complete degradation.

This heterogeneity in the system response, seemingly both oscillatory and excitable, appeared even though the same parameters were used trough the whole system, the heterogeneity being given by the cell distribution.

The appeared target centers have a range of frequencies and are stable. They also interact with each other, thus higher frequency centers dominate over the lower frequency ones. These characteristics match the observation in \textit{D.d.} patterns \cite{durston1974pacemaker}. A typical example of our simulation results can be seen in Fig. \ref{F:OneD_Spacetimes} b) and in Supplementary Video S4.

\begin{figure}[!h]
	\centering
	{\includegraphics[width=\columnwidth]{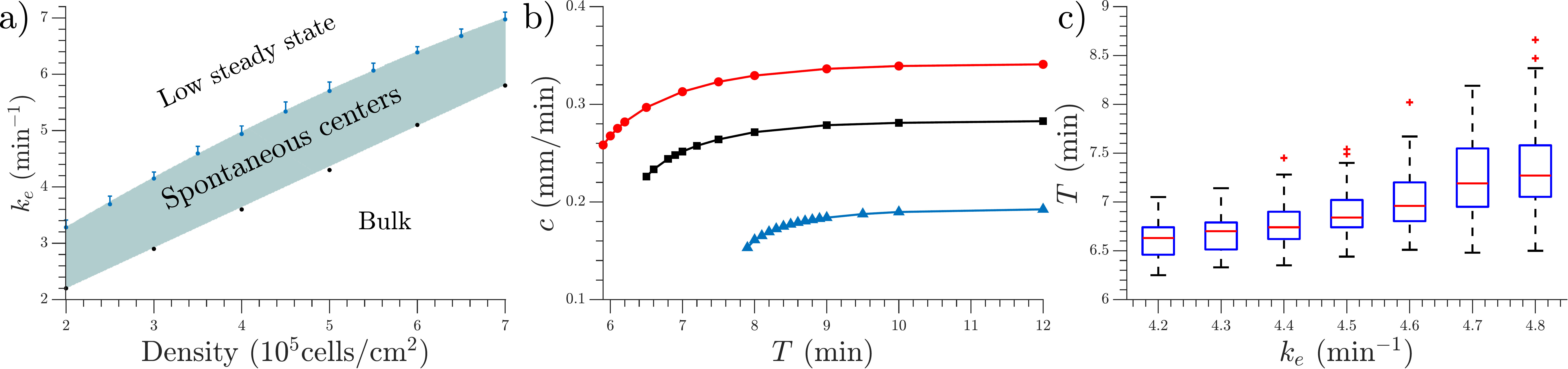}}
	\caption{ a) The shaded area indicates where spontaneous centers can be observed, depending on cell density. Blue dots indicate the degradation rate at which $75\%$ of the simulations presented spontaneous centers. Error bar extends up to $50\%$. Lower boundary marks the maximum degradation rate at which the homogeneous system presents bulk oscillations.
	b) Dispersion relations for different densities. $5\cdot10^5$ cells/cm\textsuperscript{2} (0.5 mono-layer) in red circles, $4\cdot10^5$ cells/cm\textsuperscript{2} (0.4 mono-layer) in black squares, and $3\cdot10^5$ cells/cm\textsuperscript{2} (0.3 mono-layer) in blue triangles. $k_e=5.0$ min\textsuperscript{-1}. c) Center frequency for different degradation values. 1-D simulations with density=  $4\cdot10^5$ cells/cm\textsuperscript{2}, 100 simulations used for each $k_e$ value.}
	\label{F:KeVsDensity}
\end{figure}

The range of degradation values $k_e$ in which centers can be observed  depends on cell density as is shown in Fig. \ref{F:KeVsDensity}a). Under the shaded area bulk oscillations like in Fig. \ref{F:OneD_Spacetimes} a) are observed, and above the shaded area spontaneous center do not appear, like in Fig. \ref{F:OneD_Spacetimes} c). At higher cell densities, higher degradation rates are required to observe spontaneous centers, which is consistent with the idea that phosphodiesterase is produced and released to the external media by the cells, therefore a higher concentration of cells should produce a higher concentration of phosphodiesterase and therefore a faster degradation. Nevertheless, the amount of maximum degradation does not scale linearly with density, but the size of the existence range decreases with higher density. 

The transitions between synchronized oscillations, center creation, and low steady state are not bifurcations but rather smooth and of a stochastic nature. That means, for example, that for high values of $k_e$ a small number of centers may still appear in some simulations if by chance big clusters of cells appear together. The shaded area of Fig. \ref{F:KeVsDensity} a) indicates where centers spontaneously appear in more than $75\%$ of the cases. See \textit{Methods} for a description of the boundary calculations for Fig. \ref{F:KeVsDensity} a).

To understand the nature of the traveling waves we calculated their dispersion relation, i.e. the shape of the velocity-period curve. We used the homogeneous distribution simulations mentioned in \textit{Methods} to study more controlled waves.  The waves are generated perturbing the system with a period $T$ and the velocity is calculated following each individual peak once they passed the initial transient. These dispersion relations are presented in Fig. \ref{F:KeVsDensity} b). It can be seen that they present the characteristic shape of trigger waves \cite{aliev1994dynamics}, therefore they are actual waves that produce transport of chemicals in contrast to pseudo- or phase waves which do not involve chemical transport and have a smaller dependence on diffusion. Here, diffusion is necessary to activate the next cell and with that relay the wave, another characteristic of trigger wave behaviour.   

If the homogeneous system is perturbed with higher frequencies than those depicted in Fig. \ref{F:KeVsDensity} b) it will not react with a 1:1 response, but instead it will propagate the waves with a lower frequency, usually located in the \textit{elbow} (strong curvature) area of the curve.

Furthermore, we measured the frequency of the signaling centers in 1-D simulations for a fixed cell density of $4\cdot10^5$ cells/cm\textsuperscript{2} (0.4 mono-layer) as a function of degradation rate. We measured lower center frequencies at higher degradation rates, as shown in Fig. \ref{F:KeVsDensity} c).
\begin{figure}[!h]
	{\includegraphics[width=\columnwidth]{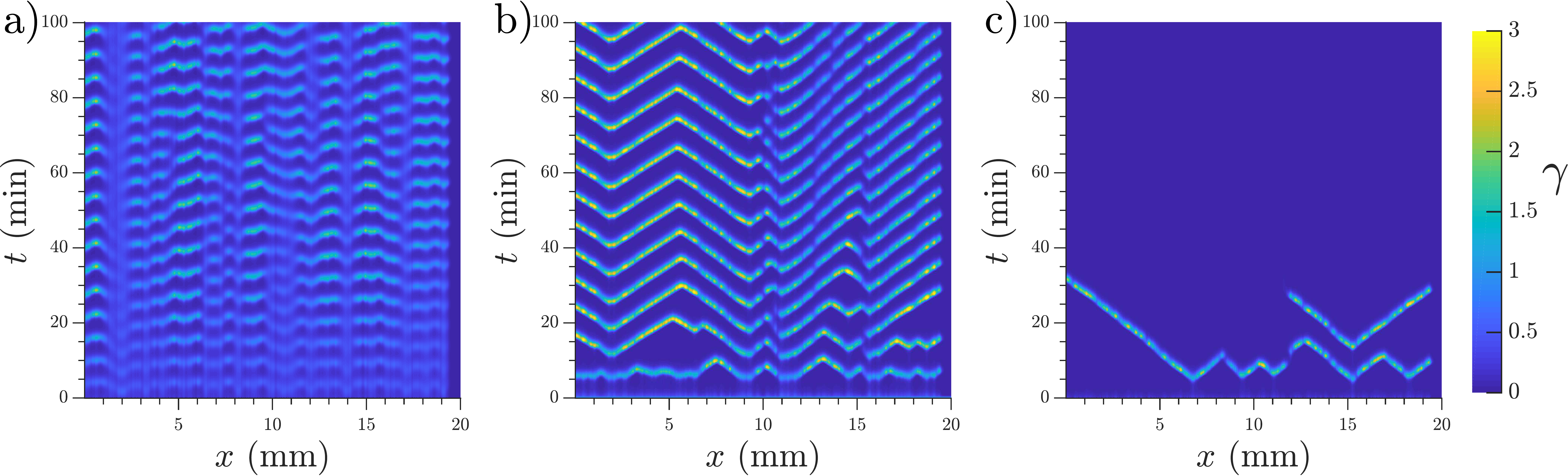}}
	\caption{{\bf Patterns at different degradation rates.} Patterns showed by the system with a random cell distribution at a density of $5\cdot10^5$ cells/cm\textsuperscript{2} (0.5 mono-layer) and different degradation rates. a) $k_e=4.0$ min\textsuperscript{-1}, b) $k_e=5.5$ min\textsuperscript{-1}, and c) $k_e=7.0$ min\textsuperscript{-1}.}
	\label{F:OneD_Spacetimes}
\end{figure}
\subsection*{Streaming}
One common test for the waves generated in a \textit{D.d.} model is if the system is capable of showing streaming once cell movement is added. Streaming is a characteristic feature of the intermediate state of \textit{D.d.} on their way towards the mounds. It consist on the alignment of the cells in a head to tail manner, thus displaying long ramified lines of higher density that spread radially from the aggregating center. 

To test this feature, we added cell movement in a simple toy-like manner. The cells would move at a constant speed $v_c$ if they sense a cAMP gradient bigger than a minimum $\partial_x\gamma>\Delta\gamma_c$ and if their percentage of active receptors is bigger than a cutoff $\rho>\rho_c$. The first rule avoids movement due to random noise and the second avoids the back-of-the-wave problem \cite{wessels1992behavior,goldstein1996traveling}. The back-of-the-wave paradox consists in taking into account that \textit{D.d.} only moves in the first half of the wave, ignoring the gradient of the decreasing cAMP in the second half which would move the cell in the opposite direction. By adding a minimum $\rho$ to allow the cells to move, they effectively move only in the first part due to the desensitization produced by the passing wave. The cells continue moving as long as these both conditions are met. Initially we did not allowed cell superposition, therefore a cell would only move towards a different space if the arriving location were empty. The updated cell positions were calculated each time step using a forward Euler scheme, after the $\gamma$, $\beta$, and $\rho$ fields were calculated. All used parameter are listed in Table \ref{t:Parameters}.

\begin{figure}[!h]
	\centering
	{\includegraphics[width=\columnwidth]{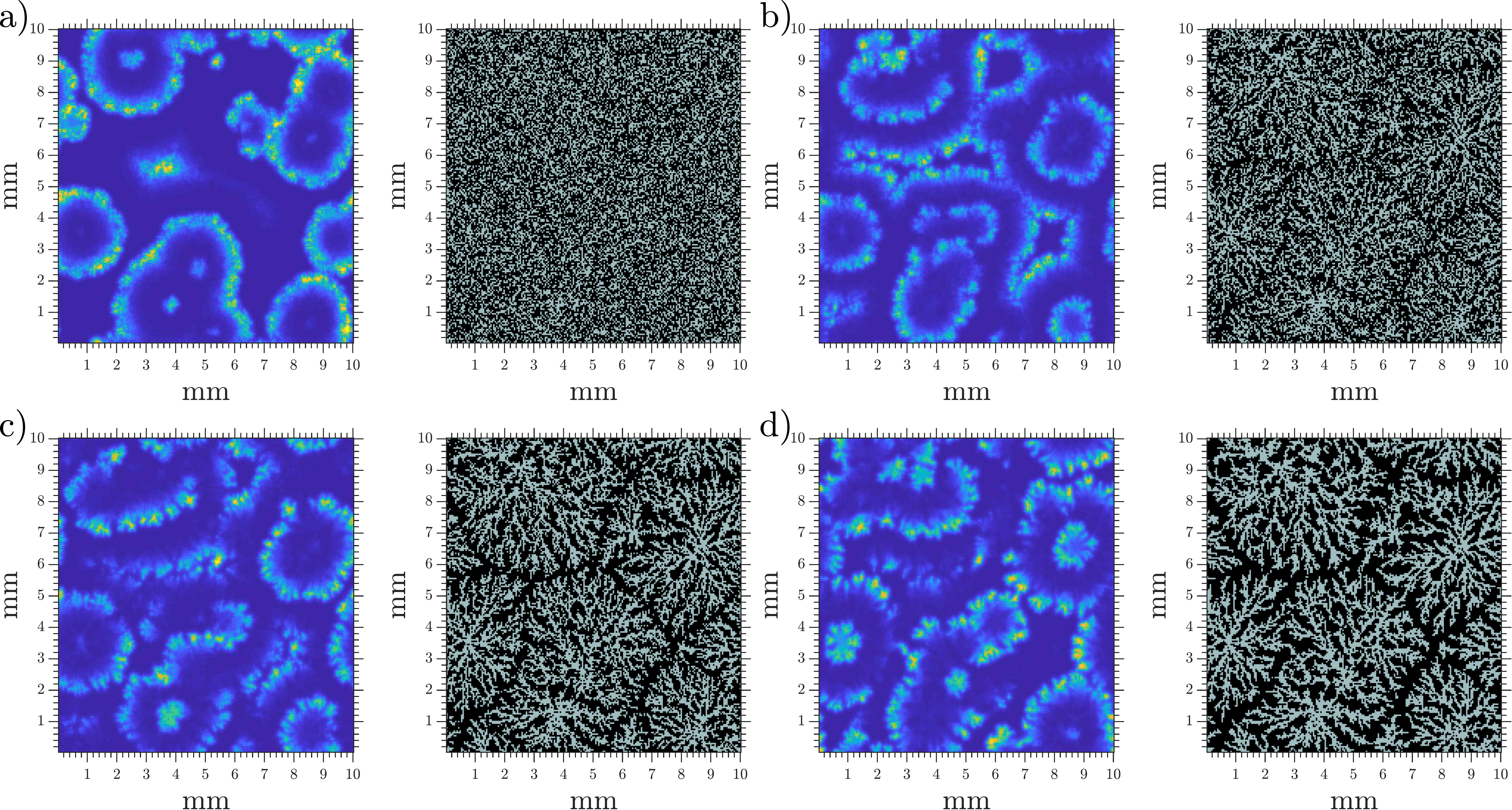}}
	\caption{ Snapshots of aggregation at different elapsed times. cAMP concentration on the left and cell distribution on the right. Same $\gamma$ scale as in Fig. \ref{F:OneD_Spacetimes}, gray squares show where cells are present and empty spaces are shown in black. Density= $4\cdot10^5$ cells/cm\textsuperscript{2} (0.4 mono-layer) and $k_e=5.7$ min\textsuperscript{-1}. a) $t=30$ min, b) $t=75$ min, c) $t=100$ min, and d) $t=150$ min.}
	\label{F:Streaming}
\end{figure}

We performed numerical simulations in 2-D where we allowed cell movement after a pattern was already established ($t\approx50$ min). Images of a typical simulation are presented in Fig. \ref{F:Streaming} where the streams can be clearly observed, thus recovering the expected behavior. The full simulation can be seen in Supplementary Video S5. It is also worth mentioning that after cells start to move some wave fronts may break and produce spirals.  

\subsection*{Cell superposition and bounded phosphodiesterase}
\begin{figure}[t]
	\centering
	{\includegraphics[width=\columnwidth]{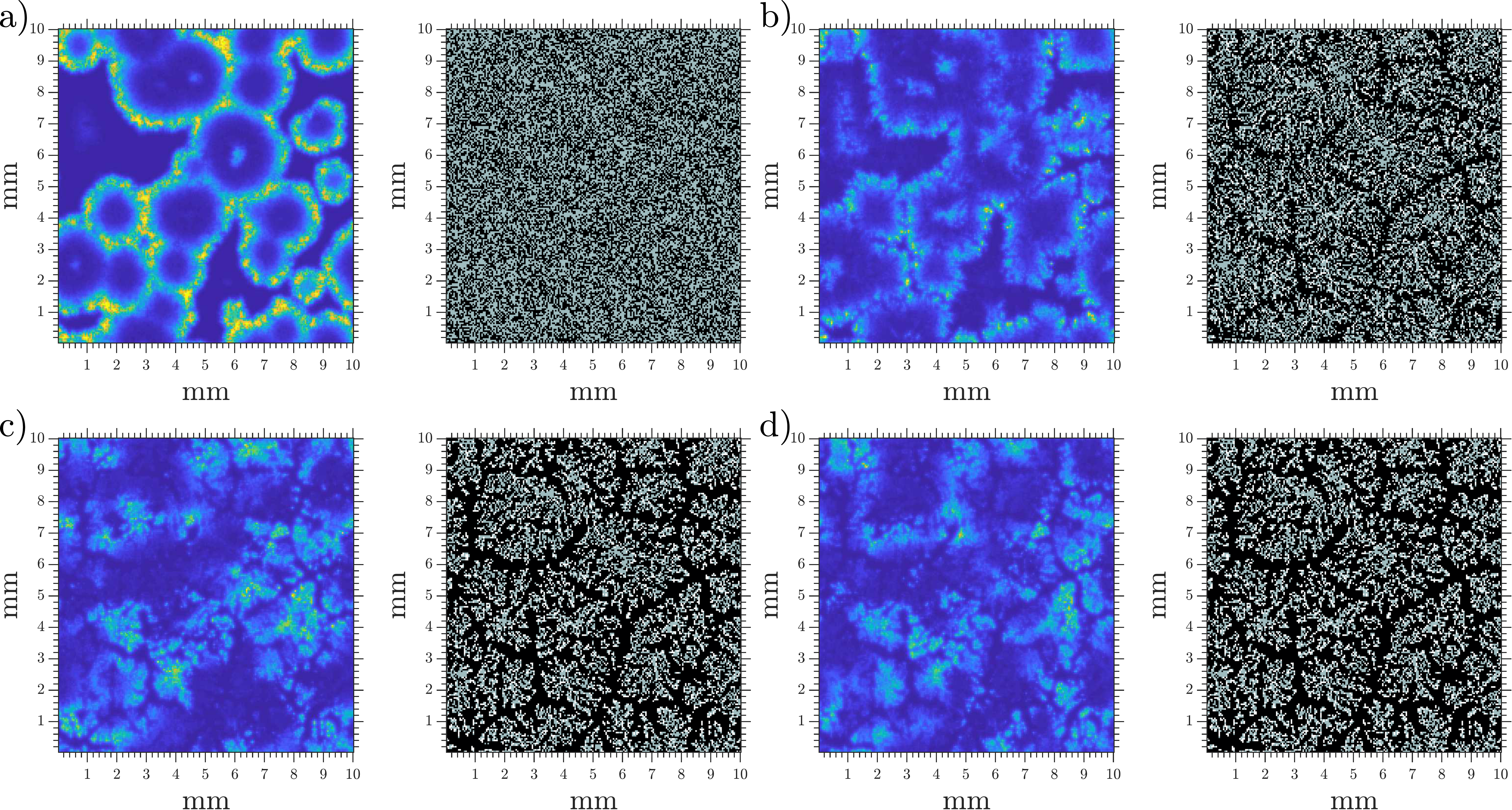}}
	\caption{Snapshots of aggregation at different elapsed times with cell superposition, but only unbounded degradation. cAMP concentration on the left and cell distribution on the right. Same $\gamma$ scale as in Fig. \ref{F:OneD_Spacetimes},  gray squares show where one cell is present, white squares contain more than one cell, and empty spaces are shown in black. Density= $5\cdot10^5$ cells/cm\textsuperscript{2} (0.5 mono-layer) and $k_e=5.7$ min\textsuperscript{-1}. a) $t=39$ min, b) $t=75$ min, c) $t=120$ min, and d) $t=150$ min.}
	\label{F:BadStreaming}
\end{figure}

As a way of perfecting our simple toy model for movement we included the possibility of cells superimposing in the same location, with a maximum of 5 cells per grid point. In order for the system to continue presenting centers and continuous cell streaming, additional degradation is needed. That is, with the previously used constant degradation the centers and aggregation streams start to break once cell clumps appear due to movement. See Fig. \ref{F:BadStreaming} in comparison to Fig. \ref{F:Streaming} where target centers and stream lines quickly break apart. A simulation showing this behaviour is presented in Supplementary Video S6.

\begin{figure}[!h]
	\centering
	{\includegraphics[width=0.6\columnwidth]{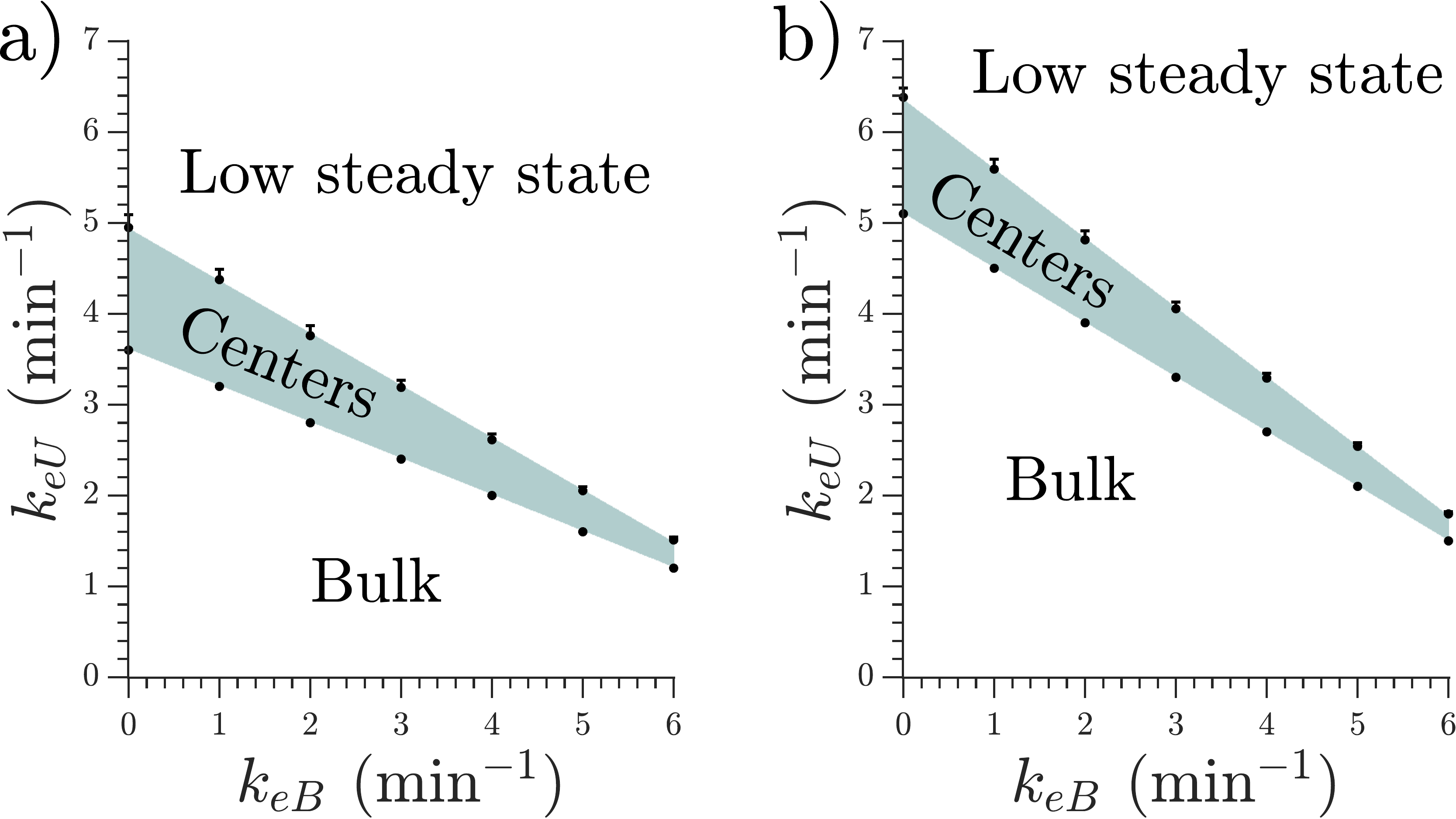}}
	\caption{ Phosphodiesterase degradation rates at which spontaneous centers exist. Below the shaded area the system presents bulk oscillations. Above the shaded area no spontaneous centers appear. Cell density is $4\cdot10^5$ cells/cm\textsuperscript{2} (0.4 mono-layer) in panel a) and  $6\cdot10^5$ cells/cm\textsuperscript{2} (0.6 mono-layer) in panel b).  Upper boundaries indicate the degradation rate at which $75\%$ of the simulations presented spontaneous centers. Error bar extends up to $50\%$. Lower boundary marks the maximum degradation rate at which the homogeneous system presents bulk oscillations. }
	\label{F:TwoKe}
\end{figure}

Many experimental studies point to the existence of phosphodiesterase bounded to the cell membrane \cite{shapiro1983comparison,bader2007seven}. Adding this extra bounded phosphodiesterase $k_{eB}$ solves these problems and the system behaves as before, producing centers which persist after movement (see Supplementary Video S7). Therefore, we conducted simulations where there is a constant degradation present on the system $k_{eU}$ which represents the unbounded, free phosphodiesterase; and a bounded degradation $k_{eB}$ which exists attached to the outside of the cell membrane and therefore exists only on the occupied spaces and in an amount dependent on the number of cells in that location. The new equations are then 

\begin{subequations}
	\begin{align}
		k_1^{-1}\partial_t\rho_i&=-f_1(\gamma(x_i,y_i))\rho_i+f_2(\gamma(x_i,y_i))(1-\rho_i),\\
		\partial_t\beta_i&=s\Phi(\rho_i,\gamma(x_i,y_i))-(k_i+k_t)\beta_i ,\\ 
		\partial_t\gamma(x,y)&=D\nabla^2\gamma(x,y)-k_{eU}\gamma(x,y) + \sum_i^N H(i,x,y) \big[k_t\beta_i/h-k_{eB}\gamma(x,y)\big].
	\end{align}
\end{subequations}
The results obtained with these equations are similar to those shown in Fig. \ref{F:OneD_Spacetimes}, that is, we distinguish three types of behaviour depending on the degradation rates. The degradation combinations for which the system still presents spontaneous centers are shown in Fig. \ref{F:TwoKe}.  Unlike in the case with no bounded phosphodiesterase, the variance of the normal distribution decreased with the bounded degradation rate, thus making the transition from centers to low steady state much sharper at higher $k_{eB}$.  

We see that there is no fixed ratio of bounded/unbounded phosphodiesterase such that the system produces centers, but rather there is a whole range of values, depending on the system density, thus allowing the cells to have some variability and still present the same behaviour. It is noteworthy that as the bounded degradation increases ($k_{eB}$), $k_{eU}$ decreases, as does the range of possible unbounded values $k_{eU}$, and the upper transition becomes sharper, since the range of values where $50-75\%$ of simulations presented centers diminishes (length of the upper errorbars). Thus the cells can have less variability in $k_{eU}$ as $k_{eB}$ increases.

\section*{Discussion}
\begin{figure}[!h]
	\centering
	{\includegraphics[width=0.8\columnwidth]{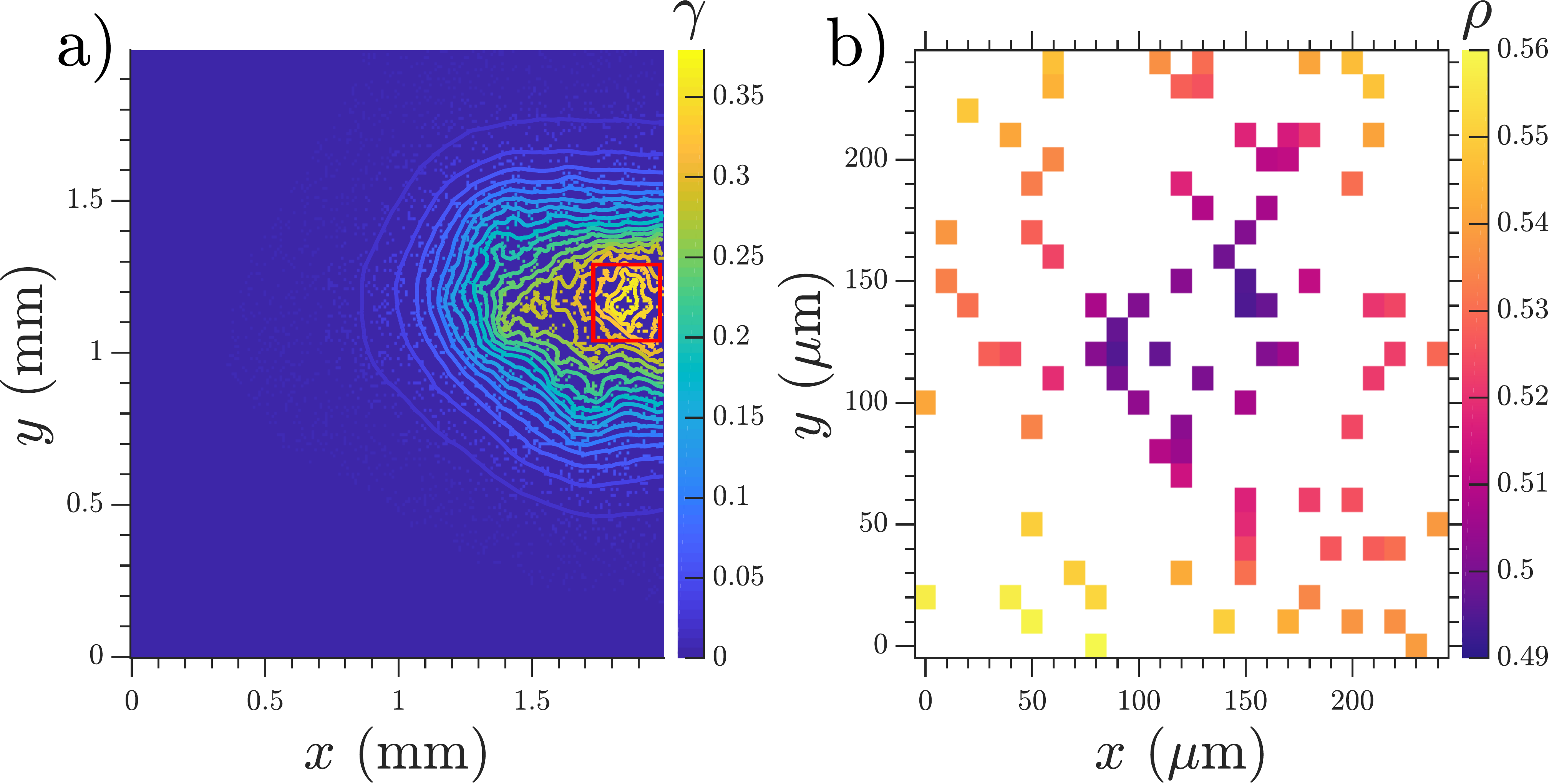}}
	\caption{Image of a numerical simulation showing a signaling center with the characteristics of those observed by Ohta et al.\cite{ohta2018red}. Density=  $1.25\cdot10^5$cells/cm\textsuperscript{2} (0.125 mono-layer), $k_e=1.56$ min\textsuperscript{-1}. a) cAMP concentration in the areas occupied by cells, with equal concentration levels of the whole extracellular media. b) Zoom of the signaling center marked with red in a) showing the percentage of active receptors $\rho$. Cells with lower $\rho$ (dark purple) are those which initially fire. Around 18 cells constitute the firing center.}
		\label{F:Com_Ohta}
\end{figure}

Over the past years several models have been used to describe the patterns observed in \textit{D.d.} Excitable models are capable of sustaining spiral and trigger waves, but in order to generate a target pattern they require either an oscillatory center or spontaneous/random firing from the cells. On the other hand, oscillatory models require some sort of inhomogeneity to be added to the system to produce stable centers and avoid bulk oscillations. It has been shown that increasing the cell parameters along a developmental path or adding random firing decides the location of the observed patterns in an artificial way \cite{geberth2008predicting,geberth2009predicting,grace2013predictability,grace2015regulation} which contrast with experimental observations showing that oscillation is a collective effect instead of the work of some specialized cells, where even the cells composing the oscillating center move continuously in and out of the signaling center \cite{gregor2010onset,ohta2018red}. We suggest that a more simple mechanism is also in play to produce centers for densities less than a mono-layer: the inhomogeneous cell distribution in the system is enough to create stable emitting centers. Our model matches the observation of Lee et al. \cite{lee1996competing} who showed that at lower densities (below mono-layer density $\approx 10^6$ cells/cm\textsuperscript{2}) target centers dominate the observed patterns, while high above confluency, spirals appear. Our simulations show that below confluency and for a fixed value of $k_e$, the number of target centers increases with cell density until the transition to bulk oscillations occur (see Fig.~\ref{F:KeVsDensity}a)). A recent experiment by Ohta et al \cite{ohta2018red} showed that at densities of the order of $1.25\cdot10^5$ cells/cm\textsuperscript{2} (0.125 mono-layer) centers can be generated by groups of roughly 13 synchronized cells in an area of $100\times100~\mu$m$^2$. These results match our observation with $k_e=1.56$ min\textsuperscript{-1} where a group of approximately 18 cells produce an oscillatory center, see Fig.~\ref{F:Com_Ohta} and Supplementary Video S8.

At higher cell densities when the pattern evolves towards spiral creation, we expect this mechanism to be less relevant, although our model did show occasionally spirals when the wave front broke due to high inhomogeneities in cell density (see Supplementary Video S9). Above mono-layer densities it has been shown that the cell development inhomogeneity \cite{lauzeral1997desynchronization,palsson1996origin} and the strength of the feedback loop on excitability \cite{sawai2005autoregulatory} play a prominent role in the spiral creation process.

We stress the simplicity of this mechanism which does not make big assumptions about the inside state of the cells, but simply requires an external degradation mechanism to be present in the media. Small groups of isolated cells reach a steady state of low cAMP concentration. We showed that this stable solution exist for small clusters, where the system has two solutions, a stable and an unstable one. Once the cluster size increases over a critical value, the cluster is no longer stable and starts to oscillate with a size dependent frequency, therefore bigger clusters have a higher frequency which allows them to dominate over smaller clusters. In simulations with randomized cell locations the centers are not necessarily large clusters of consecutive cells, but rather just areas of a higher local density, where some clusters are usually separated by small distances and act together as a big cluster. Even though this does not exactly match the scenario described in the Oscillatory Clusters section, it does provide an insight into why these centers spontaneously appear. Our model is therefore, a collective effect where groups of cells behave oscillatory and individual cells are excitable to external cAMP pulses, consistent with experimental observations \cite{gregor2010onset}, furthermore the lower cAMP level at lower densities is fundamental for their excitable behavior, displaying an oscillatory behaviour if the cAMP levels are artificially risen\cite{gregor2010onset}.

Having a constant degradation across the entire system is an oversimplification of the real setup where the cells produce different amounts of phosphodiesterase \cite{hall1993role} and it even changes over time \cite{malchow1972membrane}, but since the centers exist over a range of degradations (see Fig.~\ref{F:KeVsDensity}a)) we believe that the observed effect is robust to individual differences of phosphodiesterase production in cells, an effect which is also dampened by phosphodiesterase diffusion. Further modeling including phosphodiesterase diffusion and production is necessary to explore these possibilities.

The target patterns showed to have all the main features of those observed in experiments, like producing trigger waves, having different frequencies, and higher frequency centers taking over lower frequency ones. The system showed a range of degradation values at which it presents centers, this range decreased with increasing density, while the maximum value scaled non linearly with density. For a fixed cell density increasing the degradation rate decreased the average center frequency (see Fig.~\ref{F:KeVsDensity}c)), consistent with measurements in flow chamber experiments~\cite{gregor2010onset}.

Once cell movement was introduced, the system showed streaming when aggregating. We attribute this behaviour to the velocity dependence on density (see Fig. \ref{F:KeVsDensity} b)), as it has been shown before in similar descriptions \cite{hofer1997streaming,van1996spatial,vasiev1994simulation}. In those systems, as in ours, the cAMP wave travels faster on the areas with higher cell density thus producing a wave which is not perfectly circular. This shape leads to an aggregation which is not radially uniform, thus producing streams. In this model the velocity dependence on density comes from the wave speed difference between spaces with cells, where new cAMP is produced, and empty spaces where speed is given only by diffusion, thus on average higher density areas relay waves faster than lower density ones, producing streaming. This simple aggregation model shows that intracellular localization of cAMP or ACA is not necessary to explain streams, along with other mechanisms such as cell-cell adhesion,  chemotactical memory \cite{skoge2014cellular}, and directional rectification \cite{nakajima2014rectified}, that, although are present in the experimental system, do not seem to be fundamentally necessary for stream formation.       

The final section highlights the role of a cell bounded form of degradation. This comes into relevance once the cells have started to move and form small clumps. In order for these clumps to not disrupt the wave propagation process, a degradation that scales with density is necessary. The existence of this type of phosphodiesterase is still a matter of discussion in the experimental community\cite{garcia2009group,hall1993role,bader2007seven,xiao1997identification} and even though we believe our model provides arguments in favor of the existence of both types, the effect of locally increasing degradation can also be explained by a local accumulation of phosphodiesterase around the cells possibly due to the recent release of PDE to the extracellular media that has not yet diffused sufficiently.

In conclusion, we have presented a scenario of creation of target patterns in a model describing pattern formation in \textit{D.d.}. This scenario does not require specific stages along the cell developmental path, nor requires the introduction of random firing, both of which preselect the location of target centers. Instead, in this description wave centers appear spontaneously in areas of higher cell density. We have shown that introducing cell inhomogeneity in the Martiel-Goldbeter model creates naturally target patterns that are stable and capable of producing waves that fill the whole system. We characterized these centers, which correspond to areas of higher cell density. The size of the minimum cluster required to produce a center increases with the degradation rate of the system. If the cluster size is not big enough to sustain autonomous oscillations, it reaches a low steady state whose amplitude decreases with the degradation rate. These smaller clusters, or areas of lower local density can nevertheless be excited, thus allowing waves to propagate through the system. Therefore, this scenario reproduces the large scale organization displayed by \textit{D.d.} populations.

\section*{Methods}

The system was simulated using a fourth-order Runge-Kutta scheme with Merson correction \cite{merson1957operational} and an adaptive time step. We used finite differences and a 5 points laplacian in 2-D and 3 points in 1-D. For 1-D simulations $dx$ was selected equal to the cell size $r=10~\mu$m. To speed up calculations for most 2-D simulations, $dx=5r$ was selected, with the results being confirmed by smaller grid sizes. 

To calculate the upper boundary of Fig. \ref{F:KeVsDensity} a) and Fig. \ref{F:TwoKe} we simulated the system a hundred times for each set of parameters and recorded how many simulations presented centers. We then fitted a gaussian distribution around the boundary and from this fitting extracted the expected value for a $75\%$ of center appearance (marked as a circle) and for a $50\%$, marked as the end of the errorbar. In Fig. \ref{F:KeVsDensity}, where only unbounded phosphiesterase was present, the fitted gaussian distributions presented the same variance, differing only in their mean. In Fig. \ref{F:TwoKe} the variance lowered with increasing $k_{eB}$.

To calculate the lower boundary we used a extreme case consisting in homogeneously distributed simulations, where cells had a constant distance between them. This is the most spread out possible distribution for the cells, thus avoiding clusters as much as possible. For example, if the density is $5\cdot10^5$ cells/cm\textsuperscript{2} (0.5 mono-layer) the distribution would be one cell, then 0.01 mm of empty space, then another cell and so on. In this homogeneous system if the degradation is too small the cells show bulk oscillations. As we increased the amount of phosphodiesterase the bulk oscillations ceased and the system reached a stable steady state of low cAMP concentration. The lower boundary of Fig. \ref{F:KeVsDensity} a) and Fig.~\ref{F:TwoKe} indicates the maximum amount of phosphodiesterase such that the homogeneous system shows bulk oscillations, therefore at that degradation rate or higher any inhomogeneous cell distribution will produce target centers.  

\textbf{Data availability statement} All data used in this study will be made available upon request.
\begin{table}
\begin{center}
	\caption{\label{Table_SystemParameters}Parameters used for simulations of equation (\ref{e:MartielGoldbeter_3Comp}) in the first row. Second row: Parameters used to approximate the production function in equation (\ref{E:Approximated_production}). Third row: Parameters used when cell movement was included}
	\begin{tabular}{|c|c|c|}
		\hline
		
		$c=10$&$h=5$&$k_1=0.09$ min\textsuperscript{-1}\\
		$\kappa=18.5$ &$\sigma=0.55$ min\textsuperscript{-1}&$k_i=1.7$ min\textsuperscript{-1}\\
		$k_t=0.9$ min\textsuperscript{-1}&$\mathcal{L}_1=10$&$\mathcal{L}_2=0.005$\\
		$q=4000$& $\lambda_1=10^{-4}$&$\lambda_2=0.2575$\\
		$\theta=0.01$&$D=0.024$ mm\textsuperscript{2}/ min&$r=10$ $\mu$m\\
		\hline
		$a_0=0.04423$&$a_1=0.1526$&$a_2=328.2$\\
		\hline
		$v_c=0.02$ mm/min&$\rho_c=0.6$&$\Delta\gamma_c=25.98$ nM/mm\\
		\hline
	\end{tabular}
	\label{t:Parameters}
	\end{center}
\end{table}	 
\bibliography{bibliography_Centers}

\begin{thebibliography}{10}
\urlstyle{rm}
\expandafter\ifx\csname url\endcsname\relax
  \def\url#1{\texttt{#1}}\fi
\expandafter\ifx\csname urlprefix\endcsname\relax\def\urlprefix{URL }\fi
\expandafter\ifx\csname doiprefix\endcsname\relax\def\doiprefix{DOI: }\fi
\providecommand{\bibinfo}[2]{#2}
\providecommand{\eprint}[2][]{\url{#2}}

\bibitem{kessin2001dictyostelium}
\bibinfo{author}{Kessin, R.~H.}
\newblock \emph{\bibinfo{title}{{Dictyostelium}: evolution, cell biology, and
  the development of multicellularity}}, vol.~\bibinfo{volume}{38}
  (\bibinfo{publisher}{Cambridge University Press}, \bibinfo{year}{2001}).

\bibitem{allessie1973circus}
\bibinfo{author}{Allessie, M.~A.}, \bibinfo{author}{Bonke, F.~I.} \&
  \bibinfo{author}{Schopman, F.~J.}
\newblock \bibinfo{journal}{\bibinfo{title}{Circus movement in rabbit atrial
  muscle as a mechanism of tachycardia}}.
\newblock {\emph{\JournalTitle{Circulation research}}}
  \textbf{\bibinfo{volume}{33}}, \bibinfo{pages}{54--62},
  \doiprefix\url{10.1161/01.RES.33.1.54} (\bibinfo{year}{1973}).

\bibitem{Lechleiter123}
\bibinfo{author}{Lechleiter, J.}, \bibinfo{author}{Girard, S.},
  \bibinfo{author}{Peralta, E.} \& \bibinfo{author}{Clapham, D.}
\newblock \bibinfo{journal}{\bibinfo{title}{Spiral calcium wave propagation and
  annihilation in xenopus laevis oocytes}}.
\newblock {\emph{\JournalTitle{Science}}} \textbf{\bibinfo{volume}{252}},
  \bibinfo{pages}{123--126}, \doiprefix\url{10.1126/science.2011747}
  (\bibinfo{year}{1991}).

\bibitem{mair1996traveling}
\bibinfo{author}{Mair, T.} \& \bibinfo{author}{M{\"u}ller, S.~C.}
\newblock \bibinfo{journal}{\bibinfo{title}{Traveling nadh and proton waves
  during oscillatory glycolysis in vitro}}.
\newblock {\emph{\JournalTitle{Journal of Biological Chemistry}}}
  \textbf{\bibinfo{volume}{271}}, \bibinfo{pages}{627--630}
  (\bibinfo{year}{1996}).

\bibitem{winfree1972spiral}
\bibinfo{author}{Winfree, A.~T.}
\newblock \bibinfo{journal}{\bibinfo{title}{Spiral waves of chemical
  activity}}.
\newblock {\emph{\JournalTitle{Science}}} \textbf{\bibinfo{volume}{175}},
  \bibinfo{pages}{634--636}, \doiprefix\url{10.1126/science.175.4022.634}
  (\bibinfo{year}{1972}).

\bibitem{zaikin1970concentration}
\bibinfo{author}{Zaikin, A.} \& \bibinfo{author}{Zhabotinsky, A.}
\newblock \bibinfo{journal}{\bibinfo{title}{Concentration wave propagation in
  two-dimensional liquid-phase self-oscillating system}}.
\newblock {\emph{\JournalTitle{Nature}}} \textbf{\bibinfo{volume}{225}},
  \bibinfo{pages}{535--537}, \doiprefix\url{10.1038/225535b0}
  (\bibinfo{year}{1970}).

\bibitem{kopell1973plane}
\bibinfo{author}{Kopell, N.} \& \bibinfo{author}{Howard, L.}
\newblock \bibinfo{journal}{\bibinfo{title}{Plane wave solutions to
  reaction-diffusion equations}}.
\newblock {\emph{\JournalTitle{Studies in Applied Mathematics}}}
  \textbf{\bibinfo{volume}{52}}, \bibinfo{pages}{291--328},
  \doiprefix\url{10.1002/sapm1973524291} (\bibinfo{year}{1973}).

\bibitem{stich2002complex}
\bibinfo{author}{Stich, M.} \& \bibinfo{author}{Mikhailov, A.~S.}
\newblock \bibinfo{journal}{\bibinfo{title}{Complex pacemakers and wave sinks
  in heterogeneous oscillatory chemical systems}}.
\newblock {\emph{\JournalTitle{Zeitschrift f{\"u}r Physikalische Chemie}}}
  \textbf{\bibinfo{volume}{216}}, \bibinfo{pages}{521},
  \doiprefix\url{10.1524/zpch.2002.216.4.521} (\bibinfo{year}{2002}).

\bibitem{keener1986geometrical}
\bibinfo{author}{Keener, J.~P.}
\newblock \bibinfo{journal}{\bibinfo{title}{A geometrical theory for spiral
  waves in excitable media}}.
\newblock {\emph{\JournalTitle{SIAM Journal on Applied Mathematics}}}
  \textbf{\bibinfo{volume}{46}}, \bibinfo{pages}{1039--1056},
  \doiprefix\url{10.1137/0146062} (\bibinfo{year}{1986}).

\bibitem{tyson1988singular}
\bibinfo{author}{Tyson, J.~J.} \& \bibinfo{author}{Keener, J.~P.}
\newblock \bibinfo{journal}{\bibinfo{title}{Singular perturbation theory of
  traveling waves in excitable media (a review)}}.
\newblock {\emph{\JournalTitle{Physica D: Nonlinear Phenomena}}}
  \textbf{\bibinfo{volume}{32}}, \bibinfo{pages}{327--361},
  \doiprefix\url{10.1016/0167-2789(88)90062-0} (\bibinfo{year}{1988}).

\bibitem{levine1996positive}
\bibinfo{author}{Levine, H.}, \bibinfo{author}{Aranson, I.},
  \bibinfo{author}{Tsimring, L.} \& \bibinfo{author}{Truong, T.~V.}
\newblock \bibinfo{journal}{\bibinfo{title}{Positive genetic feedback governs
  {cAMP} spiral wave formation in {Dictyostelium}}}.
\newblock {\emph{\JournalTitle{Proceedings of the National Academy of
  Sciences}}} \textbf{\bibinfo{volume}{93}}, \bibinfo{pages}{6382--6386}
  (\bibinfo{year}{1996}).
\newblock \eprint{http://www.pnas.org/content/93/13/6382.full.pdf}.

\bibitem{noorbakhsh2015modeling}
\bibinfo{author}{Noorbakhsh, J.}, \bibinfo{author}{Schwab, D.~J.},
  \bibinfo{author}{Sgro, A.~E.}, \bibinfo{author}{Gregor, T.} \&
  \bibinfo{author}{Mehta, P.}
\newblock \bibinfo{journal}{\bibinfo{title}{Modeling oscillations and spiral
  waves in dictyostelium populations}}.
\newblock {\emph{\JournalTitle{Physical Review E}}}
  \textbf{\bibinfo{volume}{91}}, \bibinfo{pages}{062711},
  \doiprefix\url{10.1103/PhysRevE.91.062711} (\bibinfo{year}{2015}).

\bibitem{martiel1987model}
\bibinfo{author}{Martiel, J.-L.} \& \bibinfo{author}{Goldbeter, A.}
\newblock \bibinfo{journal}{\bibinfo{title}{A model based on receptor
  desensitization for cyclic {AMP} signaling in {Dictyostelium} cells}}.
\newblock {\emph{\JournalTitle{Biophysical Journal}}}
  \textbf{\bibinfo{volume}{52}}, \bibinfo{pages}{807},
  \doiprefix\url{10.1016/S0006-3495(87)83275-7} (\bibinfo{year}{1987}).

\bibitem{tyson1989spiral}
\bibinfo{author}{Tyson, J.~J.}, \bibinfo{author}{Alexander, K.~A.},
  \bibinfo{author}{Manoranjan, V.} \& \bibinfo{author}{Murray, J.}
\newblock \bibinfo{journal}{\bibinfo{title}{Spiral waves of cyclic {AMP} in a
  model of slime mold aggregation}}.
\newblock {\emph{\JournalTitle{Physica D: Nonlinear Phenomena}}}
  \textbf{\bibinfo{volume}{34}}, \bibinfo{pages}{193--207},
  \doiprefix\url{10.1016/0167-2789(89)90234-0} (\bibinfo{year}{1989}).

\bibitem{aliev1994dynamics}
\bibinfo{author}{Aliev, R.~R.} \& \bibinfo{author}{Biktashev, V.~N.}
\newblock \bibinfo{journal}{\bibinfo{title}{Dynamics of the oscillation phase
  distribution in the bz reaction}}.
\newblock {\emph{\JournalTitle{The Journal of Physical Chemistry}}}
  \textbf{\bibinfo{volume}{98}}, \bibinfo{pages}{9676--9681},
  \doiprefix\url{10.1021/j100089a049} (\bibinfo{year}{1994}).

\bibitem{hofer1997streaming}
\bibinfo{author}{H{\"o}fer, T.} \& \bibinfo{author}{Maini, P.~K.}
\newblock \bibinfo{journal}{\bibinfo{title}{Streaming instability of slime mold
  amoebae: An analytical model}}.
\newblock {\emph{\JournalTitle{Physical Review E}}}
  \textbf{\bibinfo{volume}{56}}, \bibinfo{pages}{2074},
  \doiprefix\url{10.1103/PhysRevE.56.2074} (\bibinfo{year}{1997}).

\bibitem{van1996spatial}
\bibinfo{author}{Van~Oss, C.}, \bibinfo{author}{Panfilov, A.~V.},
  \bibinfo{author}{Hogeweg, P.}, \bibinfo{author}{Siegert, F.} \&
  \bibinfo{author}{Weijer, C.~J.}
\newblock \bibinfo{journal}{\bibinfo{title}{Spatial pattern formation during
  aggregation of the slime mould dictyostelium discoideum}}.
\newblock {\emph{\JournalTitle{Journal of theoretical biology}}}
  \textbf{\bibinfo{volume}{181}}, \bibinfo{pages}{203--213},
  \doiprefix\url{10.1006/jtbi.1996.0126} (\bibinfo{year}{1996}).

\bibitem{lauzeral1997desynchronization}
\bibinfo{author}{Lauzeral, J.}, \bibinfo{author}{Halloy, J.} \&
  \bibinfo{author}{Goldbeter, A.}
\newblock \bibinfo{journal}{\bibinfo{title}{Desynchronization of cells on the
  developmental path triggers the formation of spiral waves of {cAMP} during
  {Dictyostelium} aggregation}}.
\newblock {\emph{\JournalTitle{Proceedings of the National Academy of
  Sciences}}} \textbf{\bibinfo{volume}{94}}, \bibinfo{pages}{9153--9158},
  \doiprefix\url{10.1073/pnas.94.17.9153} (\bibinfo{year}{1997}).

\bibitem{vidal2017convective}
\bibinfo{author}{Vidal-Henriquez, E.}, \bibinfo{author}{Zykov, V.},
  \bibinfo{author}{Bodenschatz, E.} \& \bibinfo{author}{Gholami, A.}
\newblock \bibinfo{journal}{\bibinfo{title}{Convective instability and boundary
  driven oscillations in a reaction-diffusion-advection model}}.
\newblock {\emph{\JournalTitle{Chaos: An Interdisciplinary Journal of Nonlinear
  Science}}} \textbf{\bibinfo{volume}{27}}, \bibinfo{pages}{103110},
  \doiprefix\url{10.1063/1.4986153} (\bibinfo{year}{2017}).

\bibitem{durston1974pacemaker}
\bibinfo{author}{Durston, A.}
\newblock \bibinfo{journal}{\bibinfo{title}{Pacemaker activity during
  aggregation in dictyostelium discoideum}}.
\newblock {\emph{\JournalTitle{Developmental biology}}}
  \textbf{\bibinfo{volume}{37}}, \bibinfo{pages}{225--235},
  \doiprefix\url{10.1016/0012-1606(74)90144-4} (\bibinfo{year}{1974}).

\bibitem{wessels1992behavior}
\bibinfo{author}{Wessels, D.}, \bibinfo{author}{Murray, J.} \&
  \bibinfo{author}{Soll, D.~R.}
\newblock \bibinfo{journal}{\bibinfo{title}{Behavior of dictyostelium amoebae
  is regulated primarily by the temporal dynamic of the natural camp wave}}.
\newblock {\emph{\JournalTitle{Cell motility and the cytoskeleton}}}
  \textbf{\bibinfo{volume}{23}}, \bibinfo{pages}{145--156},
  \doiprefix\url{10.1002/cm.970230207} (\bibinfo{year}{1992}).

\bibitem{goldstein1996traveling}
\bibinfo{author}{Goldstein, R.~E.}
\newblock \bibinfo{journal}{\bibinfo{title}{Traveling-wave chemotaxis}}.
\newblock {\emph{\JournalTitle{Physical review letters}}}
  \textbf{\bibinfo{volume}{77}}, \bibinfo{pages}{775},
  \doiprefix\url{10.1103/PhysRevLett.77.775} (\bibinfo{year}{1996}).

\bibitem{shapiro1983comparison}
\bibinfo{author}{Shapiro, R.~I.}, \bibinfo{author}{Franke, J.},
  \bibinfo{author}{Luna, E.~J.} \& \bibinfo{author}{Kessin, R.~H.}
\newblock \bibinfo{journal}{\bibinfo{title}{A comparison of the membrane-bound
  and extracellular cyclic amp phosphodiesterases of dictyostelium
  discoideum}}.
\newblock {\emph{\JournalTitle{Biochimica et Biophysica Acta (BBA)-General
  Subjects}}} \textbf{\bibinfo{volume}{758}}, \bibinfo{pages}{49--57},
  \doiprefix\url{10.1016/0304-4165(83)90009-0} (\bibinfo{year}{1983}).

\bibitem{bader2007seven}
\bibinfo{author}{Bader, S.}, \bibinfo{author}{Kortholt, A.} \&
  \bibinfo{author}{Van~Haastert, P.~J.}
\newblock \bibinfo{journal}{\bibinfo{title}{Seven dictyostelium discoideum
  phosphodiesterases degrade three pools of camp and cgmp}}.
\newblock {\emph{\JournalTitle{Biochemical Journal}}}
  \textbf{\bibinfo{volume}{402}}, \bibinfo{pages}{153--161},
  \doiprefix\url{10.1042/BJ20061153} (\bibinfo{year}{2007}).

\bibitem{geberth2008predicting}
\bibinfo{author}{Geberth, D.} \& \bibinfo{author}{H{\"u}tt, M.-T.}
\newblock \bibinfo{journal}{\bibinfo{title}{Predicting spiral wave patterns
  from cell properties in a model of biological self-organization}}.
\newblock {\emph{\JournalTitle{Physical Review E}}}
  \textbf{\bibinfo{volume}{78}}, \bibinfo{pages}{031917},
  \doiprefix\url{10.1103/PhysRevE.78.031917} (\bibinfo{year}{2008}).

\bibitem{geberth2009predicting}
\bibinfo{author}{Geberth, D.} \& \bibinfo{author}{H{\"u}tt, M.-T.}
\newblock \bibinfo{journal}{\bibinfo{title}{Predicting the distribution of
  spiral waves from cell properties in a developmental-path model of
  dictyostelium pattern formation}}.
\newblock {\emph{\JournalTitle{PLoS computational biology}}}
  \textbf{\bibinfo{volume}{5}}, \bibinfo{pages}{e1000422},
  \doiprefix\url{10.1371/journal.pcbi.1000422} (\bibinfo{year}{2009}).

\bibitem{grace2013predictability}
\bibinfo{author}{Grace, M.} \& \bibinfo{author}{H{\"u}tt, M.-T.}
\newblock \bibinfo{journal}{\bibinfo{title}{Predictability of spatio-temporal
  patterns in a lattice of coupled fitzhugh--nagumo oscillators}}.
\newblock {\emph{\JournalTitle{Journal of The Royal Society Interface}}}
  \textbf{\bibinfo{volume}{10}}, \bibinfo{pages}{20121016},
  \doiprefix\url{10.1098/rsif.2012.1016} (\bibinfo{year}{2013}).

\bibitem{grace2015regulation}
\bibinfo{author}{Grace, M.} \& \bibinfo{author}{H{\"u}tt, M.-T.}
\newblock \bibinfo{journal}{\bibinfo{title}{Regulation of spatiotemporal
  patterns by biological variability: general principles and applications to
  dictyostelium discoideum}}.
\newblock {\emph{\JournalTitle{PLoS computational biology}}}
  \textbf{\bibinfo{volume}{11}}, \bibinfo{pages}{e1004367},
  \doiprefix\url{10.1371/journal.pcbi.1004367} (\bibinfo{year}{2015}).

\bibitem{gregor2010onset}
\bibinfo{author}{Gregor, T.}, \bibinfo{author}{Fujimoto, K.},
  \bibinfo{author}{Masaki, N.} \& \bibinfo{author}{Sawai, S.}
\newblock \bibinfo{journal}{\bibinfo{title}{The onset of collective behavior in
  social amoebae}}.
\newblock {\emph{\JournalTitle{Science}}} \textbf{\bibinfo{volume}{328}},
  \bibinfo{pages}{1021--1025}, \doiprefix\url{10.1126/science.1183415}
  (\bibinfo{year}{2010}).

\bibitem{ohta2018red}
\bibinfo{author}{Ohta, Y.}, \bibinfo{author}{Furuta, T.},
  \bibinfo{author}{Nagai, T.} \& \bibinfo{author}{Horikawa, K.}
\newblock \bibinfo{journal}{\bibinfo{title}{Red fluorescent camp indicator with
  increased affinity and expanded dynamic range}}.
\newblock {\emph{\JournalTitle{Scientific reports}}}
  \textbf{\bibinfo{volume}{8}}, \bibinfo{pages}{1866},
  \doiprefix\url{10.1038/s41598-018-20251-1} (\bibinfo{year}{2018}).

\bibitem{lee1996competing}
\bibinfo{author}{Lee, K.~J.}, \bibinfo{author}{Cox, E.~C.} \&
  \bibinfo{author}{Goldstein, R.~E.}
\newblock \bibinfo{journal}{\bibinfo{title}{Competing patterns of signaling
  activity in dictyostelium discoideum}}.
\newblock {\emph{\JournalTitle{Physical review letters}}}
  \textbf{\bibinfo{volume}{76}}, \bibinfo{pages}{1174},
  \doiprefix\url{10.1103/PhysRevLett.76.1174} (\bibinfo{year}{1996}).

\bibitem{palsson1996origin}
\bibinfo{author}{P{\'a}lsson, E.} \& \bibinfo{author}{Cox, E.~C.}
\newblock \bibinfo{journal}{\bibinfo{title}{Origin and evolution of circular
  waves and spirals in dictyostelium discoideum territories}}.
\newblock {\emph{\JournalTitle{Proceedings of the National Academy of
  Sciences}}} \textbf{\bibinfo{volume}{93}}, \bibinfo{pages}{1151--1155},
  \doiprefix\url{10.1073/pnas.93.3.1151} (\bibinfo{year}{1996}).

\bibitem{sawai2005autoregulatory}
\bibinfo{author}{Sawai, S.}, \bibinfo{author}{Thomason, P.~A.} \&
  \bibinfo{author}{Cox, E.~C.}
\newblock \bibinfo{journal}{\bibinfo{title}{An autoregulatory circuit for
  long-range self-organization in dictyostelium cell populations}}.
\newblock {\emph{\JournalTitle{Nature}}} \textbf{\bibinfo{volume}{433}},
  \bibinfo{pages}{323}, \doiprefix\url{10.1038/nature03228}
  (\bibinfo{year}{2005}).

\bibitem{hall1993role}
\bibinfo{author}{Hall, A.~L.}, \bibinfo{author}{Franke, J.},
  \bibinfo{author}{Faure, M.} \& \bibinfo{author}{Kessin, R.~H.}
\newblock \bibinfo{journal}{\bibinfo{title}{The role of the cyclic nucleotide
  phosphodiesterase of {Dictyostelium} discoideum during growth, aggregation,
  and morphogenesis: overexpression and localization studies with the separate
  promoters of the pde}}.
\newblock {\emph{\JournalTitle{Developmental biology}}}
  \textbf{\bibinfo{volume}{157}}, \bibinfo{pages}{73--84},
  \doiprefix\url{10.1006/dbio.1993.1113} (\bibinfo{year}{1993}).

\bibitem{malchow1972membrane}
\bibinfo{author}{Malchow, D.}, \bibinfo{author}{N\"{a}gele, B.},
  \bibinfo{author}{Schwarz, H.} \& \bibinfo{author}{Gerisch, G.}
\newblock \bibinfo{journal}{\bibinfo{title}{Membrane-bound cyclic {AMP}
  phosphodiesterase in chemotactically responding cells of {Dictyostelium}
  discoideum}}.
\newblock {\emph{\JournalTitle{European Journal of Biochemistry}}}
  \textbf{\bibinfo{volume}{28}}, \bibinfo{pages}{136--142},
  \doiprefix\url{10.1111/j.1432-1033.1972.tb01894.x} (\bibinfo{year}{1972}).

\bibitem{vasiev1994simulation}
\bibinfo{author}{Vasiev, B.}, \bibinfo{author}{Hogeweg, P.} \&
  \bibinfo{author}{Panfilov, A.}
\newblock \bibinfo{journal}{\bibinfo{title}{Simulation of dictyostelium
  discoideum aggregation via reaction-diffusion model}}.
\newblock {\emph{\JournalTitle{Physical Review Letters}}}
  \textbf{\bibinfo{volume}{73}}, \bibinfo{pages}{3173},
  \doiprefix\url{10.1103/PhysRevLett.73.3173} (\bibinfo{year}{1994}).

\bibitem{skoge2014cellular}
\bibinfo{author}{Skoge, M.} \emph{et~al.}
\newblock \bibinfo{journal}{\bibinfo{title}{Cellular memory in eukaryotic
  chemotaxis}}.
\newblock {\emph{\JournalTitle{Proceedings of the National Academy of
  Sciences}}} \textbf{\bibinfo{volume}{111}}, \bibinfo{pages}{14448--14453},
  \doiprefix\url{10.1073/pnas.1412197111} (\bibinfo{year}{2014}).

\bibitem{nakajima2014rectified}
\bibinfo{author}{Nakajima, A.}, \bibinfo{author}{Ishihara, S.},
  \bibinfo{author}{Imoto, D.} \& \bibinfo{author}{Sawai, S.}
\newblock \bibinfo{journal}{\bibinfo{title}{Rectified directional sensing in
  long-range cell migration}}.
\newblock {\emph{\JournalTitle{Nature communications}}}
  \textbf{\bibinfo{volume}{5}}, \bibinfo{pages}{5367},
  \doiprefix\url{10.1038/ncomms6367} (\bibinfo{year}{2014}).

\bibitem{garcia2009group}
\bibinfo{author}{Garcia, G.~L.}, \bibinfo{author}{Rericha, E.~C.},
  \bibinfo{author}{Heger, C.~D.}, \bibinfo{author}{Goldsmith, P.~K.} \&
  \bibinfo{author}{Parent, C.~A.}
\newblock \bibinfo{journal}{\bibinfo{title}{The group migration of
  dictyostelium cells is regulated by extracellular chemoattractant
  degradation}}.
\newblock {\emph{\JournalTitle{Molecular biology of the cell}}}
  \textbf{\bibinfo{volume}{20}}, \bibinfo{pages}{3295--3304},
  \doiprefix\url{10.1091/mbc.e09-03-0223} (\bibinfo{year}{2009}).

\bibitem{xiao1997identification}
\bibinfo{author}{Xiao, Z.} \& \bibinfo{author}{Devreotes, P.~N.}
\newblock \bibinfo{journal}{\bibinfo{title}{Identification of
  detergent-resistant plasma membrane microdomains in dictyostelium: enrichment
  of signal transduction proteins.}}
\newblock {\emph{\JournalTitle{Molecular biology of the cell}}}
  \textbf{\bibinfo{volume}{8}}, \bibinfo{pages}{855--869},
  \doiprefix\url{10.1091/mbc.8.5.855} (\bibinfo{year}{1997}).

\bibitem{merson1957operational}
\bibinfo{author}{Merson, R.}
\newblock \bibinfo{title}{An operational method for the study of integration
  processes}.
\newblock In \emph{\bibinfo{booktitle}{Proc. Symp. Data Processing}},
  \bibinfo{pages}{1--25} (\bibinfo{year}{1957}).

\end{thebibliography}



\section*{Acknowledgements}

The authors thank E. Bodenschatz, A. Bae, and V. Zykov for fruitful discussions. E.V.H. thanks the Deutsche Akademische Austauschdienst (DAAD), Research Grants—Doctoral Programs in Germany. A.G. acknowledges MaxSynBio Consortium, which is jointly funded by the Federal Ministry of Education and Research of Germany and the Max Planck Society.

\section*{Author contributions statement}
E.V.H. and A.G designed the research, E.V.H. performed the numerical simulations and theoretical analysis, E.V.H. and A.G wrote and reviewed the manuscript.

\section*{Additional information}
 \textbf{Competing interests} The authors declare no competing interests. 

\end{document}